\documentstyle[preprint,revtex]{aps}

\begin{document}

\draft
\begin{title}
The Escape Problem for Irreversible Systems
\end{title}
\author{Robert S. Maier${}^{(1)}$ and D.~L. Stein${}^{(2)}$}
\begin{instit}
Mathematics${}^{(1)}$ and Physics${}^{(2)}$ Departments, University of
Arizona, Tucson, AZ~85721
\end{instit}

\begin{abstract}
The problem of noise-induced escape from a metastable state arises in
physics, chemistry, biology, systems engineering, and other areas.  The
problem is well understood when the underlying dynamics of the system obey
detailed balance.  When this assumption fails many of the results of
classical transition-rate theory no longer apply, and no general method
exists for computing the weak-noise asymptotics of fundamental quantities
such as the mean escape time.  In~this paper we~present a general technique
for analysing the weak-noise limit of a wide range of stochastically
perturbed continuous-time nonlinear dynamical systems.  We~simplify the
original problem, which involves solving a partial differential equation,
into one in~which only ordinary differential equations need be solved.
This allows us to resolve some old issues for the case when detailed
balance holds.  When it does not hold, we~show how the formula for the mean
escape time asymptotics depends on the dynamics of the system along the
most probable escape path.  We~also present new results on short-time
behavior and discuss the possibility of {\em focusing\/} along the escape
path.
\end{abstract}
\pacs{PACS numbers: 05.40.+j, 02.50.+s}
\narrowtext

\section{Introduction}
\label{sec:intro}
The phenomenon of escape from a locally stable equilibrium state arises in
a multitude of scientific
contexts~\cite{vanKampen81,McClintock89b,Hanggi90}.  If~a nonlinear system
is subjected to continual random perturbations (`noise'), eventually a
sufficiently large fluctuation will drive it over an intervening barrier to
a new equilibrium state.  The mean amount of time required for this to
occur typically grows exponentially as the strength of the random
perturbations tends to zero.

Research on this phenomenon has focused on the case when the nonlinear
dynamics of the system in the absence of random perturbations are specified
by a {\em potential function\/}.  In~a recent paper~\cite{MaierA} we have
introduced a new technique for computing the weak-noise asymptotics of the
mean first passage time (MFPT) to the barrier.  Our~technique, unlike the
bulk of earlier work, is not restricted to the case when the zero-noise
dynamics arise from a potential.  Because of this we~can readily and
quantitatively treat systems without `detailed balance,' whose dynamics are
determined by non-gradient drift fields, or are otherwise
time-irreversible.  We~deal here with {\em overdamped\/} systems, in~which
inertia plays no role.  Overdamped systems without detailed balance arise
in the theory of glasses and other disordered materials~\cite{Stein89},
chemical reactions far from equilibrium~\cite{Ross92}, stochastically
modelled computer networks~\cite{Nelson,Lim87,Maier7b}, evolutionary
biology~\cite{Ebeling89}, and theoretical ecology~\cite{Mangel93}.


In~the multidimensional escape problems most frequently considered in the
literature, the most probable escape path (MPEP) in the limit of weak noise
passes over a hyperbolic equilibrium point ({\em i.e.},~saddle point) of the
unperturbed dynamics.  In~the case of non-gradient drift fields exit
through an {\em unstable\/} equilibrium point can also occur~\cite{MaierA}.
Other new possibilities, such as exit through a limit
cycle~\cite{Naeh90,Talkner87}, arise as well.  However, if the unperturbed
dynamics are determined by a potential, exit in the limit of weak noise
must occur over a hyperbolic equilibrium point, and the asymptotics of the
MFPT are given by a classic formula, originally derived in the context of
chemical reactions by Eyring~\cite{Glasstone41}.

Over the years the Eyring formula has been rederived and generalized by a
variety of alternative approaches~\cite{Caroli80,Landauer61,Langer69}.
To~illustrate its use, consider a two-dimensional system whose dynamics are
specified by a sufficiently smooth drift field ${\bf u}={\bf u}(x,y)$
symmetric about the $x$-axis as displayed in Fig.~\ref{fig:field}.
$S=(x_S,0)$ and~$H=(0,0)$ denote the stable and hyperbolic equilibrium
points, and the barrier lies along the $y$-axis.  The position of a point
particle representing the system state, moving in this drift field and
subjected to additive white noise ${\bf w}(t)$, satisfies the It\^o
stochastic differential equation~\cite{Schuss80}
\begin{equation}
\label{eq:1}
dx_i(t) = u_i({\bf x}(t))\,dt +
\epsilon^{1/2}\sigma_i\,dw_i(t),\quad i=x,y.
\end{equation}
Here $\sigma_x$ and~$\sigma_y$ quantify the response of the particle to the
perturbations in the $x$~and $y$-directions; the corresponding diffusion
tensor~${\bf D}$ is ${\mathop{\rm
diag}}(D_x,D_y)={\mathop{\rm diag}}({\sigma_x}^2,{\sigma_y}^2)$
and will in general be anisotropic.  If~the drift field~${\bf u}$ is
obtained from a potential function~$\phi=\phi(x,y)$ by the formula
$u_i=-D_i\partial_i\phi$ then ${\bf u}$~will in general not be a gradient
field, but the relation between ${\bf u}$ and~$\bf D$ will ensure detailed
balance~\cite{Klosek89}.

If detailed balance holds, the Eyring formula for the $\epsilon\to0$
asymptotics of the MFPT~$\tau$ is
\begin{eqnarray}
{1\over\tau}&\sim&
\frac1\pi\sqrt
{\Big|{\partial u_x\over\partial x}(S)\Big|{\partial u_x\over\partial
x}(H)} \sqrt{(\partial u_y/\partial y)(S)\over(\partial u_y/\partial
y)(H)}\nonumber\\
&&\quad\mbox{}\times
\exp\Big[-{2\over{D_x\epsilon}}\int_{0}^{x_S}dx\ u_x(x,0)\Big]
\label{eq:2}
\end{eqnarray}
where $(\partial u_y/\partial y)(S)$ is $\partial u_y/\partial y$ evaluated
at the stable point~$S$, and $(\partial u_y/\partial y)(H)$ is $\partial
u_y/\partial y$ evaluated at the hyperbolic point~$H$, {\em etc.} In~this
example the MPEP lies along the $x$-axis; this fact was used implicitly in
the expression for the Arrhenius ({\em i.e.},~exponential) factor
in~(\ref{eq:2}).  If~the vector field were asymmetric and the MPEP were
curved, the Arrhenius factor would involve the integral $\int
({\bf{D}}^{-1}{\bf u})\cdot d{\bf x}$ taken along the
MPEP~\cite{Matkowsky88}.

Eq.~(\ref{eq:2}) displays a frequently encountered feature of weak-noise
escape problems: the dependence on~${\bf u}$ of the pre-exponential factor
in the MFPT asymptotics is limited to a dependence on the derivatives
of~${\bf u}$ near the stable and hyperbolic equilibrium points.
If~detailed balance is absent this will generally not be the
case~\cite{Landauer75,Landauer83}.  However, most quantitative work on the
case when detailed balance is absent has dealt with one-dimensional state
spaces, due to the difficulty of higher-dimensional computations.  Talkner
and H\"anggi~\cite{Talkner84} computed MFPT asymptotics for a
multidimensional model of an optically bistable system without detailed
balance, in~which the drift field did not arise from a potential.  But due
to the simplicity of their model, the prefactor did not display a
complicated dependence on the drift.  Multidimensional models without
detailed balance have not been treated in full generality.

In Sections \ref{sec:analysis} and~\ref{sec:examples} we start from scratch
and compute weak-noise MFPT asymptotics for multidimensional systems by
singular perturbation methods, using matched asymptotic expansions.
Similar approaches have been used before~\cite{Naeh90,Ludwig75}, but our
treatment yields a systematic method of computing the pre-exponential
factor, even in cases when detailed balance is absent.  As~noted, we~have
previously applied our techniques to the nonclassical case of exit over an
{\em unstable\/} equilibrium point ({\em i.e.},~an equilibrium point at
which the linearization of~${\bf u}$ has only positive
eigenvalues)~\cite{MaierA}. The MFPT asymptotics differed considerably from
the classical escape formul\ae\ such as Eq.~(\ref{eq:2}); the prefactor in
the asymptotics of~$\tau$ depended on~$\epsilon$.  In~this paper we flesh
out the ideas in our previous paper by applying them to drift fields with
the standard structure of Figure~\ref{fig:field}.  We~discover that even
when the MPEP passes over a saddle point, if~detailed balance is absent the
MFPT asymptotics likewise differ from those given by the Eyring formula.
Although the prefactor is independent of~$\epsilon$, it~depends on the
behavior of~${\bf u}$ along the entire MPEP.  We~quantify this dependence;
it~can be much more complicated than a dependence on the derivatives
of~${\bf u}$ at the endpoints.

Our treatment reveals, however, that when ${\bf u}$~is symmetric about the
MPEP as displayed in Fig.~\ref{fig:field} the prefactor can readily be
computed by integrating two coupled ordinary differential equations along
the MPEP.  Previous work has left the impression that partial differential
equations must be solved.  In~the absence of detailed balance, the MFPT
asymptotics are determined by a {\em quasipotential\/}~\cite{VF}, the
solution of a nonlinear partial differential equation.  However the
derivatives of this function along the MPEP, which turn out to determine
the prefactor, satisfy ordinary differential equations.

The prefactor is very sensitive to the behavior of the drift field in a
neighborhood of the MPEP.  Integrating the ordinary differential equations
along the MPEP requires knowledge of the extent of `differential shearing'
along the MPEP, {\em i.e.}, knowledge of the second derivatives of~${\bf
u}$ there.  In~many cases differential shearing can give rise to a {\em
focusing singularity\/} along the MPEP, in~which case more sophisticated
techniques must be used.

When detailed balance is absent, in~the limit of weak noise the short-term
dynamics of the system, as~well as the MFPT asymptotics, display novel
behavior.  As~noted, the system fluctuates around the locally stable
equilibrium state many times before it finally escapes along the MPEP.
If~the drift field~${\bf u}$ arises from a potential, each of the
unsuccessful escape attempts follows an outward `instanton' trajectory
anti-parallel to some integral curve of~${\bf u}$ and then falls back along
the same integral curve (an~`anti-instanton' trajectory).  In~the absence
of detailed balance, the time-irreversibility will generally cause the
instanton trajectories {\em not\/} to be anti-parallel to the
anti-instanton trajectories, which remain integral curves of the drift
field.  Hence unsuccessful escape attempts will proceed with high
probability along closed {\em loops\/} containing nonzero area, in~contrast
to the more familiar situation for systems with detailed balance.
We~discuss this in Section~\ref{sec:shorttime}; conclusions appear in
Section~\ref{sec:conclusions}.

\section{The Analysis}
\label{sec:analysis}
We now derive the weak-noise MFPT asymptotics for any~${\bf u}$ with the
general structure of Fig.~\ref{fig:field}, but not necessarily derived from
a potential.  We~allow anisotropic diffusion, so~long as the principal axes
of the diffusion tensor are aligned with the MPEP.  After approximating the
probability density of the system along the MPEP, we~will compute $\tau$ by
the Kramers method of computing the probability flux through the
separatrix, {\em i.e.}, the boundary of the basin of attraction of the
stable point.

The Fokker-Planck equation for the probability density $\rho$ is
$\dot\rho={\cal L}^*\rho$, with
\FL
\begin{eqnarray}
{\cal L}^* &=&
(\epsilon/2)(D_x{\partial_x}^2 + D_y{\partial_y}^2) - u_x(x,y)\partial_x
-u_y(x,y)\partial_y \nonumber\\
&&\quad - (\partial_x u_x(x,y)) - (\partial_y u_y(x,y)).
\end{eqnarray}
If~absorbing boundary conditions are imposed on the separatrix (in this
case, the $y$-axis), the MFPT in the weak-noise limit may be approximated
by $(-\lambda_1)^{-1}$, where $\lambda_1$ is the eigenvalue of the slowest
decaying density mode $\rho_1$ ({\em i.e.},~the eigenfunction of ${\cal L}^*$
whose eigenvalue has the greatest real part).  $\lambda_1$~is negative and
converges to zero as $\epsilon\to 0$, so~in the weak-noise limit
$(-\lambda_1)^{-1}$ displays Arrhenius growth.  We~have
\FL
\begin{equation}
-\lambda_1 = \int_{-\infty}^\infty
(D_x\epsilon/2)\partial_x\rho_1(0,y)\,dy\!\left/
\!\int_{0}^{\infty}\!\!\int_{-\infty}^{\infty}
\rho_1(x,y)\,dy\,dx\right.
\label{eq:3}
\end{equation}
since the right-hand side is the normalized flux of probability through the
separatrix.  The leading asymptotics of~$\lambda_1$, as~computed by this
formula, will be unaffected~\cite{Naeh90} if $\rho_1$ is taken to satisfy
${\cal L}^*\rho_1=0$.

According to~(\ref{eq:3}), to approximate $\lambda_1$ we must approximate
the normal derivative of $\rho_1$ along the separatrix.  But since the MPEP
passes through a hyperbolic point, the probability density will be
concentrated in a small region (of~size $O(\epsilon^{1/2})$) about this
point as $\epsilon\to 0$.  It~therefore suffices to approximate $\rho_1$
near the hyperbolic point.  To~evaluate the denominator of~(\ref{eq:3}),
we~must also approximate~$\rho_1$ near the stable point.  The latter
approximation is straightforward: near $(x_S,0)$ we take
\FL
\begin{eqnarray}
&&\rho_1(x,y)\label{eq:4}\\
&&\quad\sim\exp\bigl[-({D_x}^{-1}|\lambda_x(S)| (x-x_S)^2
  + {D_y}^{-1}|\lambda_y(S)| y^2)/\epsilon\bigr]\nonumber
\end{eqnarray}
where we now write $\lambda_x(S)$, $\lambda_y(S)$, $\lambda_x(H)$, and
$\lambda_y(H)$ for the (real) eigenvalues of the drift field linearized at
the stable and hyperbolic points, respectively.  This Gaussian
approximation permits the evaluation of the denominator of~(\ref{eq:3}).
It~is
$\pi\epsilon\sqrt{D_xD_y}/\sqrt{\lambda_x(S)\lambda_y(S)}+o(\epsilon)$,
since contributions from other regions are negligible.

It~remains to approximate $\rho_1$ in the vicinity of the hyperbolic point.
As a necessary first step, we approximate $\rho_1$ along the MPEP,
{\em i.e.}, the
$x$-axis.  In~the vicinity of the MPEP we use a standard WKB approximation
\begin{equation}
\rho_1(x,y)\sim K(x,y)\exp \left(-W(x,y)/\epsilon\right).
\label{eq:6}
\end{equation}
Here $W$ is the quasipotential, the `nonequilibrium potential' investigated
by Graham and others~\cite{Graham89,Kupferman92}.  $K$~satisfies a transport
equation, and $W$~an eikonal (Hamilton-Jacobi) equation: ${H({\bf x},\nabla
W)=0}$, with $H$ the Wentzell-Freidlin Hamiltonian~\cite{VF}
\begin{equation}
H({\bf x},{\bf p})={{D_x}\over2}{p_x}^2+{{D_y}\over2}{p_y}^2+
{\bf u}({\bf x})\cdot {\bf p}.
\label{eq:7}
\end{equation}
So~$W(x,y)$ can be viewed as the classical action of the zero-energy
trajectory from $(x_S,0)$ to~$(x,y)$.  If~detailed balance holds, {\em
i.e.}, $u_i=-D_i\partial_i\phi$ for some potential function~$\phi$, it~is
easily checked that $W(x,y)=2\phi(x,y)+C$, for $C$ an~arbitrary constant.
If~detailed balance is absent then $W$~will in general be more difficult to
compute.  The zero-energy classical trajectories determined by the
Hamiltonian of~(\ref{eq:7}) include the instanton trajectories; we~will return
to this point in Section~\ref{sec:shorttime}.

Given the drift field structure shown in Fig.~\ref{fig:field}, and assuming
sufficient smoothness, we~can expand ${\bf u}$ near the $x$-axis in powers
of~$y$:
\begin{eqnarray}
\label{eq:ux}
u_x &=&v_0(x) + v_2(x)y^2 + O(y^4) \\
u_y &=&u_1(x)y + O(y^3).
\label{eq:uy}
\end{eqnarray}
Using this notation, we have, {\em e.g.}, $\lambda_y(S)=u_1(x_S)$ and
$\lambda_x(H)=\partial v_0/\partial x(0)$.  The function~$v_2$ measures the
extent of differential shearing along the MPEP.

Again assuming sufficient smoothness, we~can expand $W$ itself near the
$x$-axis in powers of~$y$:
\begin{equation}
W(x,y)=f_0(x)+f_2(x)y^2+O(y^4)
\label{eq:8}
\end{equation}
where $f_2(x)$ is a measure of the transverse behavior of the `WKB~tube' of
probability current along the MPEP; clearly,
$f_2(x)=\frac12\partial^2W/\partial y^2(x,0)$.  Physically, $f_2$~measures
the transverse lengthscale on which probability density is nonnegligible;
the tube profile will be approximately Gaussian, with variance proportional
to~$\epsilon/f_2(x)$ at position~$x$.

Substituting the {\em Ansatz\/}~(\ref{eq:6}) into ${\cal L}^*\rho_1=0$ and
equating the coefficients of powers of~$\epsilon$ yields a system of
equations for the the functions $f_0$, $f_2$,~$K$ in~terms of $v_0$,
$v_2$,~$u_1$.  (The system closes: no~higher derivatives of~${\bf u}$ or~$W$
enter.)  The first of these is the one-dimensional eikonal equation
$H(x,f_0'(x))=0$, where
\begin{equation}
H(x,p_x)=\frac{D_x}2{p_x}^2+v_0(x)p_x
\label{eq:9}
\end{equation}
is a Hamiltonian governing motion along the MPEP.
For this Hamiltonian,
\begin{equation}
\dot x = D_x p_x + v_0(x)
\end{equation}
is the expression relating velocity and momentum.

By the eikonal equation, $f_0'(x)$ can be viewed as the momentum~$p_x$ of an
on-axis classical trajectory with zero energy.  It~follows from the
Hamiltonian~(\ref{eq:9}) that there are only two solutions to the
eikonal equation:
$f_0'(x)=-2v_0(x)/D_x$ (arising from an instanton trajectory moving against
the drift, with $\dot x=-v_0(x)$), and $f_0\equiv0$ (arising from an
anti-instanton trajectory following the drift, with $\dot x=+v_0(x)$).
That these trajectories are directly anti-parallel to each other is a
consequence of our assumption that the drift field is symmetric about the
$x$-axis MPEP.  Since physical exit trajectories begin at the stable point
and terminate near the hyperbolic point, when constructing the WKB
approximation~(\ref{eq:6}) we use the instanton solution
$f_0'(x)=-2v_0(x)/D_x$ rather than
the anti-instanton solution $f_0\equiv0$.

The equation for $f_2$ is a nonlinear Riccati equation, and may be written as
\begin{equation}
\dot f_2 = -2D_y{f_2}^2 -2u_1f_2 + 2v_0 v_2/D_x
\label{eq:10}
\end{equation}
where the dot signifies derivative with respect to instanton transit time.
(Since the instanton trajectory satisfies $\dot x=-v_0(x)$, the left-hand
side equals $-v_0{f_2}'$.)  This equation is new, and has an immediate
physical interpretation: it~describes how the WKB tube spreads out or
contracts under the influence of its environment, as~one progresses along
the MPEP.  The final inhomogeneous term on the right-hand side
of~(\ref{eq:10}) is particularly important.  If~the differential
shearing~$v_2$ is sufficiently negative in some portion of the MPEP, this
term can drive $f_2$ to zero before the hyperbolic point is
reached~\cite{MaierA,Maier7b}.  If~this occurs, to~leading order the WKB
tube splays out to infinite width.  The tube width actually remains finite,
but becomes larger than $O(\epsilon^{1/2})$; this becomes clear if
higher-order terms are taken into account.

If $f_2$ goes negative, further integration of~(\ref{eq:10}) will normally
drive $f_2$ to~$-\infty$ in finite time; for a pictorial example of this
`focusing' phenomenon, which is really the appearance of a singularity in
the nonequilibrium potential, see Fig.~1 of Day~\cite{Day87}.
Eq.~(\ref{eq:10}) accordingly gives a new quantitative measure of the
validity of our WKB approximation: for~it to be valid, $f_2$~must remain
positive along the entire MPEP.

It~has not been recognized that the validity of the WKB approximation along
a straight MPEP can be so easily checked.  Nonlinear equations
resembling~(\ref{eq:10}) have been derived by Schuss~\cite{Schuss80} and
Talkner and Ryter~\cite{Talkner87,Talkner83}, but in the context of motion
along the separatrix rather than along the MPEP.  Their equations are
homogeneous rather than inhomogeneous, so they give rise to no
singularities.  Ludwig and Mangel~\cite{Ludwig75,Mangel77} derived a
homogeneous equation in the context of motion along an anti-instanton
trajectory, {\em i.e.}, motion following the drift.

The third and final equation is for the function~$K$.  The transport
equation for $K(x,y)$ is a well-known partial differential
equation~\cite{Naeh90,Talkner87}, but for the computation of the MFPT
asymptotics an~ordinary differential equation along the MPEP will suffice.
We~may approximate~$K$ within the WKB tube as a function of~$x$ alone,
in~which case we find
\begin{equation}
\dot K/K =-u_1 - D_y f_2
\label{eq:11}
\end{equation}
In the vicinity of the MPEP, the WKB approximation~(\ref{eq:6}) arising from
perturbations of the on-axis instanton trajectory $\dot x=-v_0(x)$ is
completely determined by the ordinary differential equations (\ref{eq:10})
and~(\ref{eq:11}), together with the initial conditions
$f_2(x_S)=|\lambda_y(S)|/D_y$ and $K(x_S)=1$.  These follow from the
requirement that the WKB approximation match the Gaussian
approximation~(\ref{eq:4}) near the stable point~$(x_S,0)$.

Finally we can approximate $\rho_1$ near the hyperbolic point $H=(0,0)$.
In~the diffusion-dominated region within an $O(\epsilon^{1/2})$ distance of
the origin, $\rho_1$~satisfies the equation
\FL
\begin{equation}
(\epsilon/2)(D_x\partial_x^2\rho_1 + D_y\partial_y^2\rho_1) -
\partial_x(\lambda_xx\rho_1) -\partial_y(\lambda_yy\rho_1)=0
\label{eq:12}
\end{equation}
which is separable: $\rho_1(x,y)=\rho_1^x(x)\rho_1^y(y)$.  By~the symmetry
of~${\bf u}$, $\rho_1^y$~must be even; the absorbing boundary condition on the
separatrix implies that $\rho_1^x$~must be odd.  The separation constant,
and the overall normalization, are found by requiring that this solution
match the WKB approximation~(\ref{eq:6}).

For the matching to occur, the separation constant must equal zero.  The
symmetry requirements mandate that
\begin{equation}
\label{eq:13}
\rho_1(x,y) \sim
C e^{X^2/4} y_2(\case{1}/{2},X)\,
\exp \left(-\vert\lambda_y(H)\vert y^2/D_y\epsilon\right).
\end{equation}
We~have introduced the rescaled variable
$X=x\sqrt{2\lambda_x(H)/D_x\epsilon}$,
and $y_2(\frac12,\cdot)$ is the odd parabolic cylinder
function~\cite{Abramowitz65} of index~$\frac12$.  $y_2(\frac12,\cdot)$~can
be expressed in terms of elementary functions, but we write $\rho_1$ in
terms of~it to facilitate comparison with our earlier work on exit over an
unstable equilibrium point~\cite{MaierA}.  There a similar expression
occurred, but the index of the parabolic cylinder function was not fixed.
Cf.~the treatment of Caroli {\em et al.}~\cite{Caroli80}, in~whose
treatment Weber functions ({\em i.e.},~parabolic cylinder functions) of
variable
index were used.  Note that $y_2'(\frac12,0)=1$ by~definition, and that
$y_2(\frac12,X)\sim\sqrt{\pi/2}\,e^{X^2/4}$ as~$X\to+\infty$.

The normalization constant~$C$ is obtained by matching the $X\to+\infty$
asymptotics of $\rho_1(x,y)$ to the WKB solution near the hyperbolic point.
This gives
\begin{equation}
C=K(0)\sqrt{2/\pi}\, \exp\left[ -\frac{2}{D_x\epsilon}\int_0^{x_S}
u_0(x)\,dx\right]
\label{eq:14}
\end{equation}
where $K(0)$ must be computed by integrating~(\ref{eq:11}) along the
MPEP from $S$ to~$H$.  Since $f_2$ appears on the right-hand side
of~(\ref{eq:11}), this in turn requires an integration of~(\ref{eq:10})
along the MPEP.

The essential point here is that the behavior of $\rho_1$ near the
hyperbolic point --- in~particular, its overall normalization --- can in
general only be obtained by integrating along the MPEP, and will depend on
the entire history of~${\bf u}$ and its first and second partial derivatives
along~it.  Substituting our approximations (\ref{eq:4}) and~(\ref{eq:13})
for~$\rho_1$ into Eq.~(\ref{eq:3}) now yields
\begin{eqnarray}
\frac1\tau&\sim&  {{K(0)}\over{\pi}} \sqrt{\vert\lambda_x(S)\vert\lambda_x(H)}
\sqrt{{{\lambda_y(S)}\over{\lambda_y(H)}}}\nonumber\\
&& \qquad\quad\mbox{}\times\exp \left[-{2\over{D_x\epsilon}}\int_0^{x_S}
u_0(x)\,dx\right]
\label{eq:15}
\end{eqnarray}
which resembles the Eyring formula~(\ref{eq:2}) but contains a new
`frequency factor'~$K(0)$.


Formul\ae\ similar to Eq.~(\ref{eq:15}) appear in Schuss~\cite{Schuss80}
and Talkner~\cite{Talkner87}.  However, Eq.~(\ref{eq:15}) does not indicate
how $\rho_0(H)$ is to be found.  Our treatment makes it clear that for
drift fields with the structure of Fig.~\ref{fig:field}, the frequency
factor~$K(0)$ and the extent to which the MFPT asymptotics differ from the
Eyring formula are most easily computed by integrating the ordinary
differential equations (\ref{eq:10}) and~(\ref{eq:11}) along the MPEP.

\section{Explicit Examples}
\label{sec:examples}
If the drift field is derived from a potential, we recover the Eyring
formula as follows.  As~noted, if $u_i = -D_i\partial_i\phi$ for some
potential function~$\phi$ then $W(x,y)=2\phi(x,y)$ up to an additive
constant.  This simplifies the calculation of $f_2$: the nonlinear Riccati
equation for~$f_2$ need not be integrated explicitly, though it could~be.
Necessarily
\begin{eqnarray}
f_2(x)&=&\left.{1\over2}{\partial^2 W(x,y)\over\partial y^2}\right|_{(x,0)}=
\left.{\partial^2 \phi(x,y)\over\partial y^2}\right|_{(x,0)} \nonumber\\
&=& \left. -{D_y}^{-1}{\partial u_y\over\partial y}\right|_{(x,0)}=-u_1(x)/D_y.
\label{eq:17}
\end{eqnarray}
Eq.~(\ref{eq:17}) settles a recurring issue.  It~is sometimes
assumed~\cite{Caroli80}, in~studying multidimensional escapes, that if
${\bf u}$ varies too rapidly along the MPEP the WKB approximation may break
down due to $f_2$ going negative at some point along the MPEP.  We~have
just shown that if detailed balance holds, {\em this will never happen\/},
so long as $u_1$ satisfies the obvious transverse stability condition of
being strictly negative.  In~the language of chemical physics, that
$f_2(x)=-u_1(x)/D_y$ for every~$x$ says that in this case, transverse
fluctuations are in local thermal equilibrium at all points~$x$ along the
MPEP.  If~detailed balance is lacking, $f_2(x)$ will not depend on $u_1(x)$
alone; Eq.~(\ref{eq:10}) makes this observation quantitative.

It follows immediately from (\ref{eq:11}) and~(\ref{eq:17}) that
$K\equiv{\rm{constant}}$; since $K(x_S)=1$, $K(0)=1$ also.  So~the
formula~(\ref{eq:15}) reduces to the Eyring formula when ${\bf u}$~is
derived from a potential.

To illustrate the power of our technique, we~now examine an overdamped
system without detailed balance where the frequency factor~$K(0)$ cannot be
computed analytically.  Suppose for simplicity that $D_x=D_y=1$, and
consider the drift field
\begin{eqnarray}
\label{eq:20a}
u_x&=&x-x^3-\alpha xy^2 \\
u_y&=&-y-x^2y.
\label{eq:20}
\end{eqnarray}
It~is easily checked that this has the structure shown in
Fig.~\ref{fig:field} for any~$\alpha$, with $x_S=1$, and that for
$\alpha=1$ (and only $\alpha=1$) this~${\bf u}$ is derivable from a
potential~$\phi$.  In~this case
\begin{equation}
\phi(x,y)=-{\textstyle\frac12}x^2+
{\textstyle\frac14}x^4+{\textstyle\frac12}y^2+{\textstyle\frac12}
x^2y^2.
\label{eq:21}
\end{equation}
Note that for all~$\alpha$, the eigenvalues $\lambda_x(S)=-2$,
$\lambda_y(S)=-2$, $\lambda_x(H)=1$, and $\lambda_y(H)=-1$.

If $\alpha=0$, $u_x$ is independent of~$y$ and the escape time problem
becomes essentially one-dimensional. In~particular, the MFPT asymptotics
are given by the one-dimensional Kramers formula~\cite{Hanggi90}
\begin{equation}
\frac1\tau \sim \frac1{\pi}
\sqrt{\vert\lambda_x(S)\vert\lambda_x(H)}
\,\exp \left(-\frac{2}{\epsilon} \Delta \phi\right).
\label{eq:Kramers}
\end{equation}
As~$\alpha$ varies between zero and one, we expect the asymptotics
of~$\tau$ to interpolate smoothly between those of the Kramers
formula~(\ref{eq:Kramers}) and those of the Eyring formula~(\ref{eq:2}).

Our technique can be used to solve for $K(0)$ and the prefactor in the MFPT
asymptotics for any value of~$\alpha$.  In~general, a numerical integration
of the equations (\ref{eq:10}) and~(\ref{eq:11}) for $f_2$ and~$K$ is
required.  Though this is straightforward, it is more illustrative of the
value of our technique to perturb about the values of~$\alpha$ at which
analytic solutions can be obtained.

We therefore consider the drift field (\ref{eq:20a}),(\ref{eq:20}) with
$\alpha=1+\delta$, $|\delta|\ll 1$, and solve for the first-order
(in~$\delta$) correction to $K(0)$ in~(\ref{eq:15}).  We~have
$u_1=-(1+x^2)$, $v_0=x-x^3$, and $v_2=-(1+\delta)x$.  Assuming sufficient
smoothness, we~expand $f_2$ in~powers of~$\delta$:
\begin{eqnarray}
\nonumber
f_2 &=& f_2^{(0)}+\delta f_2^{(1)}+ o(\delta) \\
& =&(1+x^2)+\delta f_2^{(1)}+ o(\delta)
\label{eq:22}
\end{eqnarray}
since $f_2^{(0)} = -u_1/D_y$ when $\alpha=1$.  Substituting this expansion
into~(\ref{eq:10}), we find a {\em linear\/} equation for the first-order
correction~$f_2^{(1)}$:
\begin{equation}
\frac{d}{dx}{f_2^{(1)}}=2\Big({1+x^2\over x-x^3}\Big)f_2^{(1)}+2x
\label{eq:23}
\end{equation}
This inhomogeneous first-order equation is easily solved:
\begin{equation}
f_2^{(1)}=
{2x^2\over
(1-x^2)^2}\Bigl[\log{x}+(1-x^2)+{\textstyle\frac14}(x^4-1)\Bigr].
\label{eq:24}
\end{equation}
Notice that when $\delta\neq0$ and detailed balance is lacking, to~first
order $f_2(x)$~will differ from~$-u_1(x)/D_y$ at all~$x$ along the MPEP.
So~if $\alpha\neq1$, we~expect that the transverse fluctuations are in
local thermal equilibrium nowhere along the MPEP.

Using Eq.~(11), and converting the derivative with respect to instanton
transit time to a space derivative gives the equation for~$K$:
\begin{equation}
K'/K=\delta{2x\over(1-x^2)^3}\Bigl(\log{x}+{\textstyle\frac34}+
{\textstyle\frac14}x^4-x^2\Bigr).
\label{eq:25}
\end{equation}
Its solution is
\FL
\begin{eqnarray}
&&K(x)/K(1)\label{eq:26}\\
&&\quad=\exp\Bigl[-\delta\Big(1-x^2-(2x^4-4x^2)\log{x}\Big)/4(1-x^2)^2\Bigr].
\nonumber\end{eqnarray} From this we find, to order~$\delta$,
$K(0)=K(1)(1+{\textstyle\frac38}\delta)=1+{\textstyle\frac38}\delta$.  So
\FL
\begin{equation}
\frac1\tau\sim
\frac{1+{\textstyle\frac38}\delta}{\pi}
\sqrt{|\lambda_x(S)|\lambda_x(H)}\,
\sqrt{\frac{\lambda_y(S)}{\lambda_y(H)}}
\,\exp \left(-\frac{2}{\epsilon} \Delta \phi\right)
\label{eq:corrected}
\end{equation}
to first order in~$\delta$.  Eq.~(\ref{eq:corrected}) displays the
first-order correction to the Eyring asymptotics.

Just as we have perturbed about the analytically soluble $\alpha=1$ model,
we~could also perturb about the analytically soluble $\alpha=0$ model,
where detailed balance is altogether absent.  When $\alpha=0$ the
nonequilibrium potential $W(x,y)$ cannot be computed explicitly.  However
its transverse second derivative~$f_2$ along the MPEP can be computed
explicitly by integrating~(\ref{eq:10}).  The computation of the
first-order correction~$f_2^{(1)}$ yields corrections to the Kramers
asymptotics of~(\ref{eq:Kramers}); details are left to the reader.

Notice that since $v_2=-\alpha x$, when $\alpha$~is sufficiently large and
positive the function~$f_2$ obtained by integrating the differential
equation~(\ref{eq:10}) along the MPEP will develop a zero at some point
between $x=1$ and~$x=0$.  At~this point the WKB tube will (naively, see
Section~\ref{sec:analysis})
splay out to infinite width, and the WKB approximation will break
down.  This is qualitatively reasonable: by~(\ref{eq:20a}), if~$\alpha>0$
the differential shearing near the MPEP is such as to enhance the
likelihood of motion away from the MPEP in order to overcome the resistance
of the drift field.

\section{Short Time Behavior and Unsuccessful Escapes}
\label{sec:shorttime}
We turn to the short-term dynamics of our model, and consider the
trajectories followed by the particle in the exponentially many
unsuccessful escape attempts which precede the final successful escape
along the MPEP.

The situation in the presence of detailed balance is well known.
An~unsuccessful escape attempt, defined as a sample path which leaves the
diffusion-dominated region of size~$O(\epsilon^{1/2})$ surrounding the
stable point~$S$, with high probability follows a trajectory moving
anti-parallel to the drift.  The return path follows a deterministic
trajectory satisfying $\dot{\bf x}={\bf u}({\bf x})$.  This symmetry is a
consequence of time-reversibility.

If detailed balance is absent, the instanton trajectories ({\em i.e.},~the
classical trajectories with nonzero momentum and zero energy, as~determined
by the Wentzell-Freidlin Hamiltonian, emanating from the stable point)
serve as the most likely unsuccessful escape paths.  This can be seen in
several ways.  The Wentzell-Freidlin Hamiltonian~(\ref{eq:7}) is dual to
the Onsager-Machlup Lagrangian
\begin{equation}
L({\bf x},\dot{\bf x})= \frac1{2D_x} |\dot x - u_x|^2
+\frac1{2D_y} |\dot y - u_y|^2.
\end{equation}
The probability that a sample path will be close to any specified
trajectory ${\bf x}^*(t)$,\ $0\le t\le T$, should to leading order fall
off exponentially in~$\epsilon$, with rate constant equal to the integral
of the Lagrangian along~${\bf x}^*$.  This integral is minimized when
${\bf x}^*$~is a {\em classical\/} trajectory, and it is easily seen that
minimizing the integral over transit time~$T$,
for fixed endpoints ${\bf x}^*(0)$ and~${\bf x}^*(T)$,
selects out the classical
trajectories of zero energy.

\stepcounter{footnote}
There are in general many such trajectories, including the deterministic
trajectories which simply follow the drift.  (For~them, $L$~is identically
zero.)  We~have adopted the field-theoretic nomenclature of
Coleman~\cite{Coleman79}, according to which the most likely exiting
trajectories are called `instantons,' and the return trajectories are
anti-instantons.
A~careful treatment, using functional integral
methods, shows that the most likely trajectories,
{\em i.e.}, the local minima of the  action functional,  consist of
any number of (pieces~of) instanton trajectories, each followed by the
corresponding anti-instanton trajectory: the relaxation toward the stable
point~$S$ provided by the deterministic dynamics.  Our nonequilibrium
potential $W(x,y)$ is simply the classical action of the instanton
trajectory terminating at~$(x,y)$.

Graham and others~\cite{Graham89} have stressed the importance of the
nonequilibrium potential in determining the long-term dynamics or MFPT,
as~well as the MPEP.  However its importance in short-term dynamics has
seldom if ever been clearly enunciated.  {\em If~the system is observed to
be in some internal state~$S'$ far from the stable state~$S$, with high
probability it reached~$S'$ by following an instanton trajectory, and will
return to~$S$ by following an anti-instanton trajectory.} This applies to
states~$S'$ which are outside the diffusion-dominated zone of
size~$O(\epsilon^{1/2})$ surrounding~$S$, and the phrase ``with high
probability'' can be given a quantitative meaning in the weak-noise limit.
In the absence of detailed balance the outward instanton trajectories are
not anti-parallel to the inward anti-instanton trajectories, and the
resulting closed loops contain nonzero area.  This is a consequence of
time-irreversibility.


Fig.~\ref{fig:trajectories} illustrates this phenomenon.  If $D_x=D_y=1$
and the drift field~${\bf u}$ is given by (\ref{eq:20a}) and~(\ref{eq:20}) with
$\alpha=0$, detailed balance is absent.  Though the nonequilibrium
potential cannot be computed explicitly except along the $x$-axis, the
instanton and anti-instanton trajectories are easily computed numerically.
Typical trajectories are shown.  That the $x$-axis instanton trajectory
({\em i.e.},~the MPEP) is atypical,
being directly anti-parallel to the drift, is
in part a consequence of the symmetry of our model.  It~is also a
consequence of the fact that by (\ref{eq:ux}) and~(\ref{eq:uy}), we~are
assuming $\partial u_y/\partial x = \partial u_x/\partial y$ to hold along
the MPEP.  Within the WKB tube of width~$O(\epsilon^{1/2})$ about the MPEP,
to~high accuracy ${\bf u}$~is irrotational and detailed balance holds.  However
the small deviations from detailed balance due to differential shearing
can, as~we have seen, prevent transverse fluctuations from being in local
thermal equilibrium.

Fig.~\ref{fig:trajectories} also illustrates the phenomenon of focusing:
several of the instanton trajectories displayed intersect each other,
though not along the MPEP.  Due~to such intersections (which typically form
{\em caustics\/}~\cite{Schulman,Chinarov93}, or~singular surfaces of
codimension unity) the computation of the nonequilibrium potential~$W$ will
in general require a minimization over trajectories which are {\em
piecewise classical\/} (with zero energy) rather than classical.  Some
preliminary investigations of this effect have been made by Graham and
T\'el~\cite{Graham86}, but it is not yet clear how to handle foci appearing
along the MPEP.  In~our model it is easily checked that $f_2$~being driven
through zero to~$-\infty$ is the sign of a focus along the MPEP, in~fact of
a cusp catastrophe~\cite{Schulman}.  The appropriate extensions to the WKB
approximation are now under study.

Note that though our instanton and anti-instanton trajectories are incident
on the stable point~$S$ and/or the hyperbolic point~$H$, the portion of any
trajectory within an $O(\epsilon^{1/2})$ distance of either equilibrium
point is without physical significance.  On~this lengthscale diffusion
dominates in the sense that the particle can diffuse from any point to any
other within $O(1)$~time.

This sheds light on the seeming paradox of the trajectories having infinite
transit time.  Since $\dot{\bf x}={\bf u}({\bf x})$ for anti-instanton
trajectories an infinite amount of time is needed to reach~$S$; similarly
an infinite amount of time is needed for the instanton trajectories to
emerge from~$S$.  The physical escape trajectories, both successful and
unsuccessful, should be viewed as emerging from the diffusion-dominated
zone rather than from~$S$ itself.  So~for example any unsuccessful escape
trajectory directed along the $x$-axis will have transit time
\begin{equation}
|\lambda_x(S)|^{-1} \log (1/\epsilon^{1/2}) + O(1).
\end{equation}
The transit time of the return path, {\em i.e.},
the corresponding anti-instanton
trajectory, will be the same.  Similarly the amount of time needed for the
final, successful escape trajectory to approach~$H$ will be
\begin{equation}
\lambda_x(H)^{-1} \log (1/\epsilon^{1/2}) + O(1).
\end{equation}
In~all
\begin{equation}
[|\lambda_x(S)|^{-1} + \lambda_x(H)^{-1}] \log (1/\epsilon^{1/2}) + O(1)
\end{equation}
time units will be required for the MPEP to be traversed in full during the
successful escape.

We~see that there are two timescales: the exponentially large MFPT and the
logarithmic timescale on which escape attempts occur.  In~the weak-noise
limit the latter is very brief compared to the former; the number of
unsuccessful escape attempts grows exponentially.  The successful escape,
when it finally occurs, will on the MFPT timescale occur almost
instantaneously; this justifies the term `instanton.'

\section{Summary and Conclusions}
\label{sec:conclusions}
We have seen that even in the absence of detailed balance, the weak-noise
MFPT asymptotics of a nonlinear system with a point attractor can readily
be computed.  There is no simple expression for the pre-exponential factor
in the asymptotics, such as that given by the Eyring formula.  Rather,
it~follows by integrating the differential equations (\ref{eq:10})
and~(\ref{eq:11}) along the MPEP.  This technique clarifies the new
phenomena that can occur, such as the formation of focusing singularities
due to differential shearing along the MPEP.  The effects of these
singularities on the MFPT asymptotics will be treated in a future
publication.

Our treatment makes it clear that the time-irreversibility present in
systems without detailed balance manifests itself even unto the shortest
time scales.  It~is well known that in~such systems stationary states, and
the quasi-stationary state we~employ to compute the MFPT asymptotics, are
characterized by a global circulation of probability.  However the
circulation discussed in Section~\ref{sec:shorttime} differs from the
conventional sort~\cite{Landauer88} in that it occurs locally, at the level
of {\em individual\/} unsuccessful escape attempts.  Recognition of this
point is essential for a proper computation of the MFPT asymptotics.

In this paper we have treated the two-dimensional case on account of its
simplicity, but our treatment generalizes to higher dimensions.  Curved
MPEPs, diffusion tensors whose principal axes are not aligned with the
MPEP, and nonconstant diffusion tensors can all be treated by appropriate
extensions of the techniques presented here.  In~higher-dimensional models,
or models with curved MPEPs, the Riccati equation~(\ref{eq:10}) must be
replaced by a {\em matrix\/} Riccati equation.  This extension is related
to a standard result of stochastic control theory~\cite{Fleming75}: the
effects of perturbing about optimal trajectories, obtained by solving a
Hamilton-Jacobi equation, can be computed by solving a matrix Riccati
equation.

Most studies of the overdamped escape problem have dealt with
time-reversible stochastic models, with detailed balance.  This is a
reflection of the origins of the escape problem in statistical mechanics,
where nonlinear dynamics are usually derived from a potential.  In~broader
applications of stochastic modelling, which include engineering as well as
the biological and social sciences, there is little reason to expect
time-reversibility.  Our approach should prove useful in the wider context.

\acknowledgements
The research of the first author was partially supported by the National
Science Foundation under grant NCR-90-16211.  This research was conducted
in part while the authors were in~residence at the Complex Systems Summer
School in Santa~F\'e.

\figure{A~drift field ${\bf u}$ symmetric about the $x$-axis,
with stable equilibrium point~$S=(x_S,0)$ and hyperbolic point~$H=(0,0)$.
$\lambda_x(S)$~and~$\lambda_y(S)$, the eigenvalues of the linearization
of~${\bf u}$ at~$S$, are taken to be real and negative.  This sketch assumes
$\lambda_x(S)=\lambda_y(S)$, but that is not assumed in the text.
\label{fig:field}}

\figure{Typical instanton and anti-instanton
trajectories for the drift field specified by equations
(\protect\ref{eq:20a}) and~(\protect\ref{eq:20}), with $\alpha=0$.  The two
families are superimposed: the anti-instanton trajectories are largely
straight, but the instanton trajectories curve strongly to the right.  The
focus mentioned in the text occurs to the right.  Diffusion here is
isotropic; $D_x=D_y=1$.\label{fig:trajectories}}

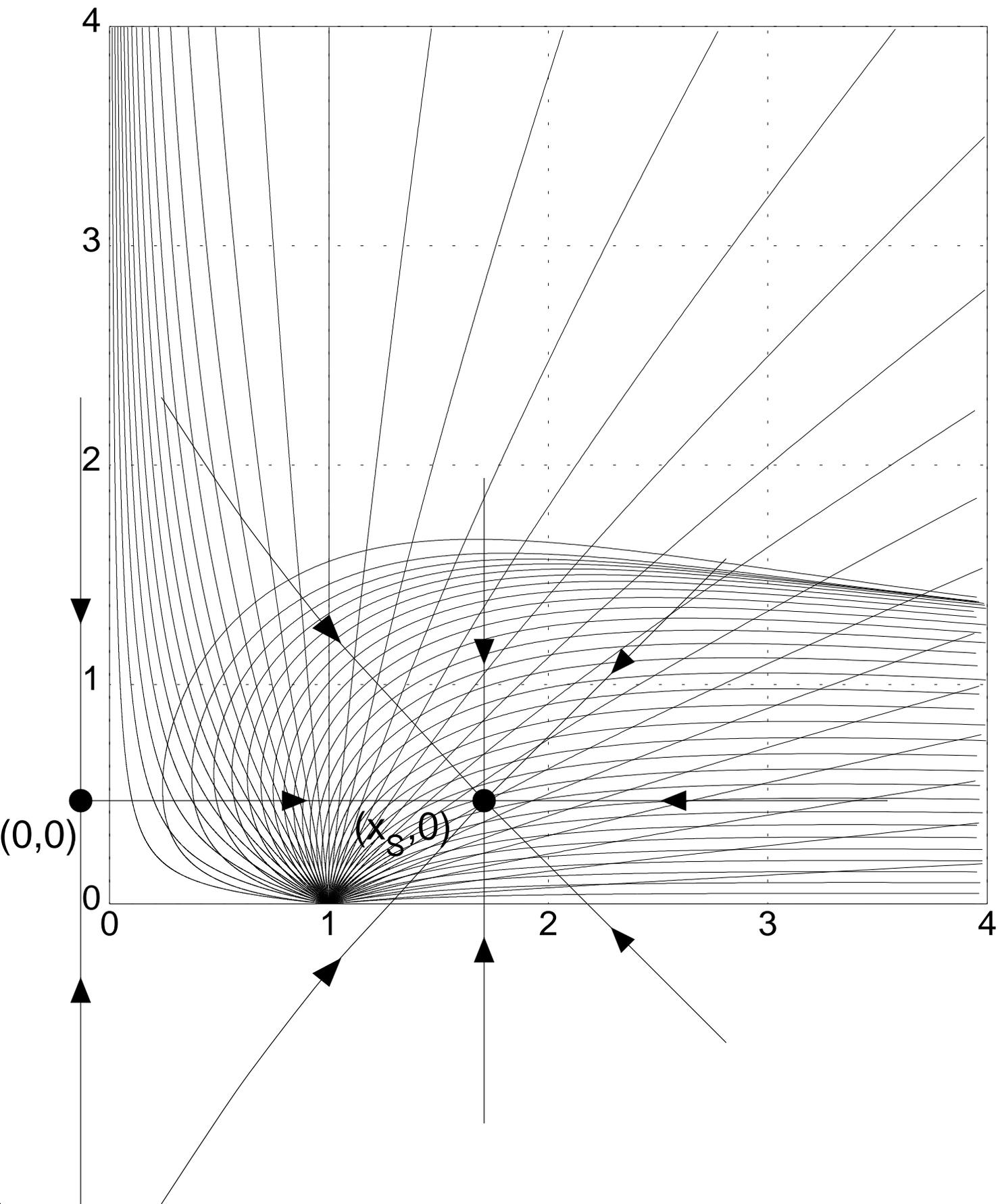
\end{document}